\documentclass[aps,prb,preprint,superscriptaddress]{revtex4}

\usepackage[dvips]{graphicx}
\usepackage{amsmath}

\begin{document}
\title{Minimum Anisotropy of a Magnetic Nanoparticle out of Equilibrium}
\author{W. Jiang}
\affiliation{School of Physics, Georgia Institute of Technology, 837 State Street, Atlanta, Georgia 30332, USA}

\author{P. Gartland}
 \email{pgartland3@gatech.edu}
 \affiliation{School of Physics, Georgia Institute of Technology, 837 State Street, Atlanta, Georgia 30332, USA}

 \author{D. Davidovi\'c}%
 \email{dragomir.davidovic@physics.gatech.edu.}
\affiliation{School of Physics, Georgia Institute of Technology, 837 State Street, Atlanta, Georgia 30332, USA}

\date{\today}


\begin{abstract}

In this article we study magnetotransport in single nanoparticles of Ni, Py=Ni$_{0.8}$Fe$_{0.2}$, Co, and Fe, with volumes $15\pm 6$nm$^3$, using sequential electron tunneling at 4.2K temperature.  We measure current versus magnetic field in the ensembles of nominally the same samples, and obtain the abundances of magnetic hysteresis. The hysteresis abundance  varies among the metals as Ni:Py:Co:Fe=4\,:50\,:100\,:100(\%), in good correlation with the magnetostatic and magnetocrystalline anisotropy.
The abrupt change in the hysteresis abundance among these metals suggests a concept of minimum magnetic anisotropy required for magnetic hysteresis, which is found to be $\approx 13$meV. The minimum anisotropy is explained in terms of the residual magnetization noise arising from the spin-orbit torques generated by sequential electron tunneling. The magnetic hysteresis abundances are weakly dependent on the tunneling current through the nanoparticle, which we attribute to
current dependent damping.
%
\end{abstract}


\maketitle

Magnetic anisotropy in ferromagnets is vital in magneto-electronic applications, such as giant magnetoresitance~\cite{fert,grunberg} and spin-transfer torque.~\cite{slonczewski1,berger,ralph2} For example, in some applications, a strong spin-orbit anisotropy is desired in order to establish a hard or fixed reference magnetic layer, while in other applications, it is beneficial to use a weaker anisotropy in order to fabricate an easily manipulated, soft or free magnetic layer. The ability to tune the degree of anisotropy for various applications is therefore of utmost importance. In thermal equilibrium, the minimum anisotropy necessary for magnetic hysteresis is temperature dependent~\cite{neel,brown,wernsdorfer}.
In this article, we address the minimum anisotropy in the case of a voltage-biased metallic ferromagnetic nanoparticle First studies of discrete levels and magnetic hysteresis in metallic ferromagnetic nanoparticles have been done on Co nanoparticles~\cite{gueron,deshmukh,jiang1}. Here we discuss the magnetic hysteresis abundances in single electron tunneling devices containing similarly sized single nanoparticles
of Ni, Py=Ni$_{0.8}$Fe$_{0.2}$, Co, and Fe. At 4.2K temperature. the probability that a given nanoparticle sample will display magnetic hysteresis in current versus magnetic field,
at any bias voltage, was found to vary as follows: Ni:Py:Co:Fe=$0.04$\,:$0.5$\,:$1$\,:$1$. 
The very small (high) probability of magnetic hysteresis in the Ni (Fe and Co) nanoparticles suggests a concept of minimum magnetic anisotropy necessary for magnetic hysteresis, comparable to the average magnetic anisotropy of Py nanoparticle. The minimum magnetic anisotropy is explained here in terms of the fluctuating spin-orbit torques exerted on the magnetization by sequential electron tunneling. These torques lead to the saturation of the effective magnetic temperature at low temperatures.
In order for the nanoparticle to exhibit magnetic hysteresis at 4.2K, the blocking temperature must be larger than the residual temperature. 
The magnetic hysteresis abundances are found to be independent of the tunneling current through the sample, which suggests that the damping is proportional to the tunneling current.
\section{Experiment}

\begin{figure}[htp]
\includegraphics[width=0.95\textwidth]{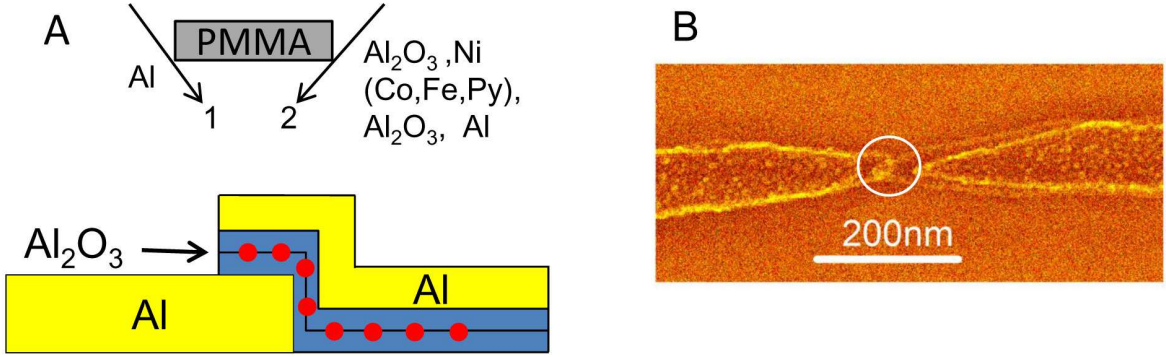}
    \caption{A: Sketch showing the sample fabrication process. B: Scanning electron micrograph of a
typical sample.}
    \label{fig1}
\end{figure}
Our samples consist of similarly-sized ferromagnetic nanoparticles tunnel-coupled to two Al
leads via amorphous aluminum oxide barriers. First, a polymethilmetachryllate bridge
is defined by electron-beam lithography on a SiO$_2$ substrate using a technique developed
previously, as sketched in Fig.~1A. Next, we deposit $10$nm of Al along direction 1. Then,
we switch the deposition to direction 2, and deposit $1.5$nm of Al$_2$O$_3$, $0.5$-$1.2$nm of ferromagnetic material, 1.5nm of Al$_2$O$_3$, topped off by $10$nm of Al, followed by liftoff in acetone. The
tunnel junction is formed by the small overlap between the two leads as shown in the circled
part in Fig.~1B. The nanoparticles are embedded in the matrix of Al$_2$O$_3$ in the overlap.
The nominal thickness of deposited Co, Ni, and Py is $0.5$-$0.6$nm. At that thickness, the
deposited metals form isolated nanoparticles approximately $1-5$nm in diameter. We find that if
we deposit Fe at the nominally thickness $0.6$nm, then the resulting samples are generally
insulating. Thus, the deposited Fe thickness is increased to $1-1.2$nm, which yields samples
in the same resistance range as in samples of Co, Ni, and Py. We suppose that because Fe
can be easily oxidized, Fe nanoparticles are surrounded by iron oxide shells. Thus, our sample
characterization suggests that it is appropriate to attribute the difference among Co, Ni, and
Py samples to intrinsic material effects rather than size discrepancies, while, in Fe nanoparticles,
the comparison is complicated by the uncertainty in the size of the metallic core. Still, we find the comparison with Fe to be fair, because of the wide range of nanoparticle diameters
involved.

\begin{figure}[htp]
\includegraphics[width=0.6\textwidth]{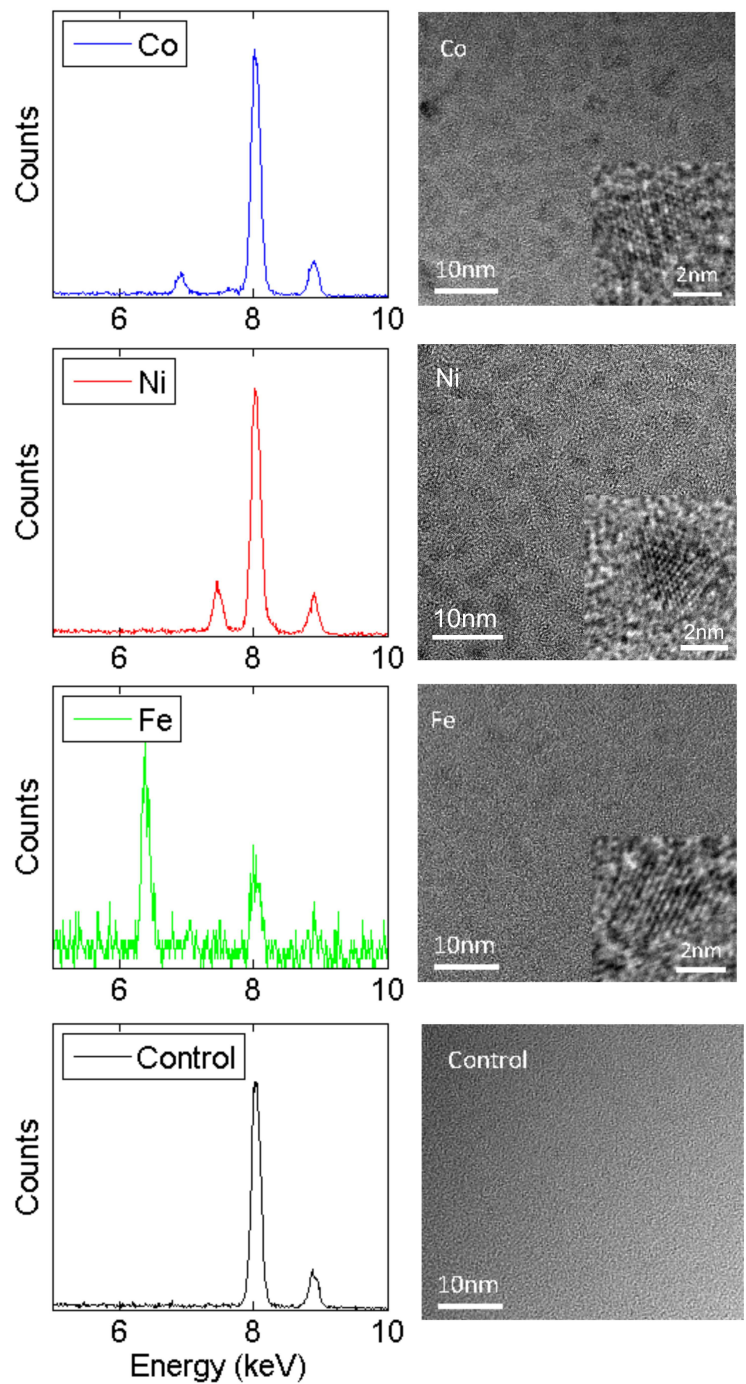}
    \caption{EDS spectra and TEM images of aluminum oxide surface topped with 0.6nm of Co (top row), 0.6nm of Ni (second row), 1.2nm of Fe (third row), and control Al$_2$O$_3$
     on TEM grid (bottom row).}
    \label{fig2}
\end{figure}
We obtain the transmission electron microscope (TEM) image of the deposited pure
aluminum oxide surface, and the aluminum oxide surface topped with nominally 1.2nm
of Fe, $0.55$nm of Ni, and $0.6$nm of Co, as shown in Fig.~2. The deposition is done
immediately prior to loading the sample in the TEM. Pure aluminum oxide surface appears
completely amorphous, with no visible signs of crystalline structure. In comparison, single
crystal structure can be identified in the TEM images for Fe, Ni, and Co. From the TEM
image, the areal coverage of Ni nanoparticles is $44$\% and the nanoparticle density is $3.6\cdot 10^4\mu$m$^{-2}$.
Assuming that the nanoparticles have pancake shape, the average area and the height of the
particles are $0.44/3.6\cdot 10^4\mu$m$^{-2}\approx 12$nm$^2$ and $0.55$nm$/0.44 = 1.25$nm, respectively. The
standard deviation of the nanoparticle's area is $40$\% of the average area, which is estimated by
the shape analysis in $120$ Ni neighboring nanoparticles. Thus, the volume of the Ni nanoparticles is
$15\pm 6$nm$^3$, where $6$nm$^3$ is the standard deviation. The volume distribution in Co nanoparticles
is similar to that in Ni.
We take the energy dispersive x-ray spectroscopy (EDS) for the samples we imaged, which
confirms the materials deposited on the aluminum oxide surface.

\section{Measurements}

\begin{figure}[htp]
\includegraphics[width=0.8\textwidth]{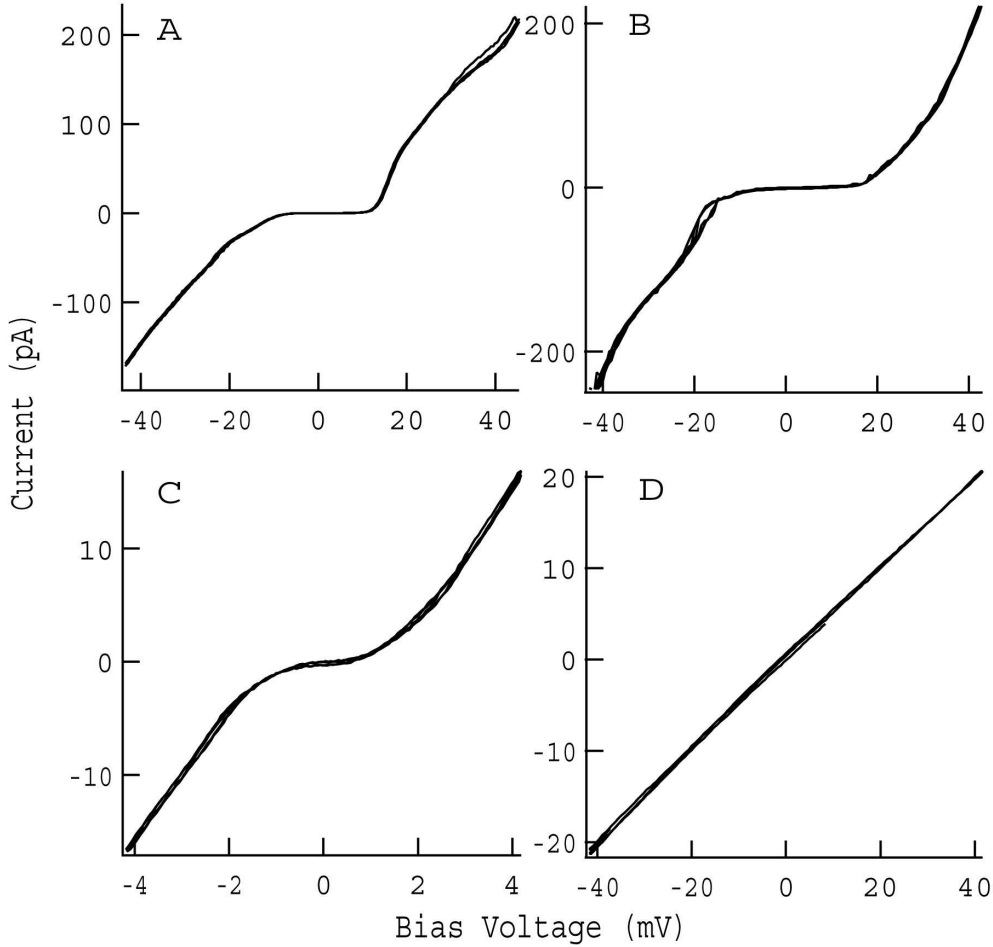}
    \caption{Current versus voltage in four representative samples at 4.2K temperature.A.Co, B.Ni, C.Fe nanoparticles, and D.Leaky pure Al$_2$O$_3$
tunneling junctions. }
    \label{fig3}
\end{figure}

The IV curves are measured using an Ithaco model 1211 current preamplifier and are reproducible with voltage sweeps. Figures~\ref{fig3} A, B, and C display the IV curves of three representative tunneling junctions with embedded nanoparticles of Co, Fe, and Ni, respectively, in samples immersed in liquid Helium at 4.2K. The IV curves display Coulomb blockade (CB) which confirms electron tunneling via metallic nanoparticles, but no discrete levels are resolved. We also measure tunneling junctions containing only the aluminum
oxide, without embedded metallic nanoparticles. Those junctions are generally
insulating but some are not, e.g., there may be leakage. The IV curves of those leaky
junctions are linear, as shown in Fig.~\ref{fig3}D, as expected for simple tunnel junctions. We show the IV curve in a pure aluminum oxide junction, to demonstrate that the CB in the samples with embedded metallic nanoparticles originate from tunneling via those nanoparticles. The issue here is that, as will be shown immediately below, some of the samples with embedded nanoparticles do not display any hysteresis in current versus magnetic field at 4.2K. Since those samples also display Coulomb blockade in the I-V curve, we can conclude that the absence of hysteresis is intrinsic to the nanoparticles, rather than  an artifact from tunneling through a possibly leaky aluminum oxide.

\begin{figure}[htp]
\includegraphics[width=0.8\textwidth]{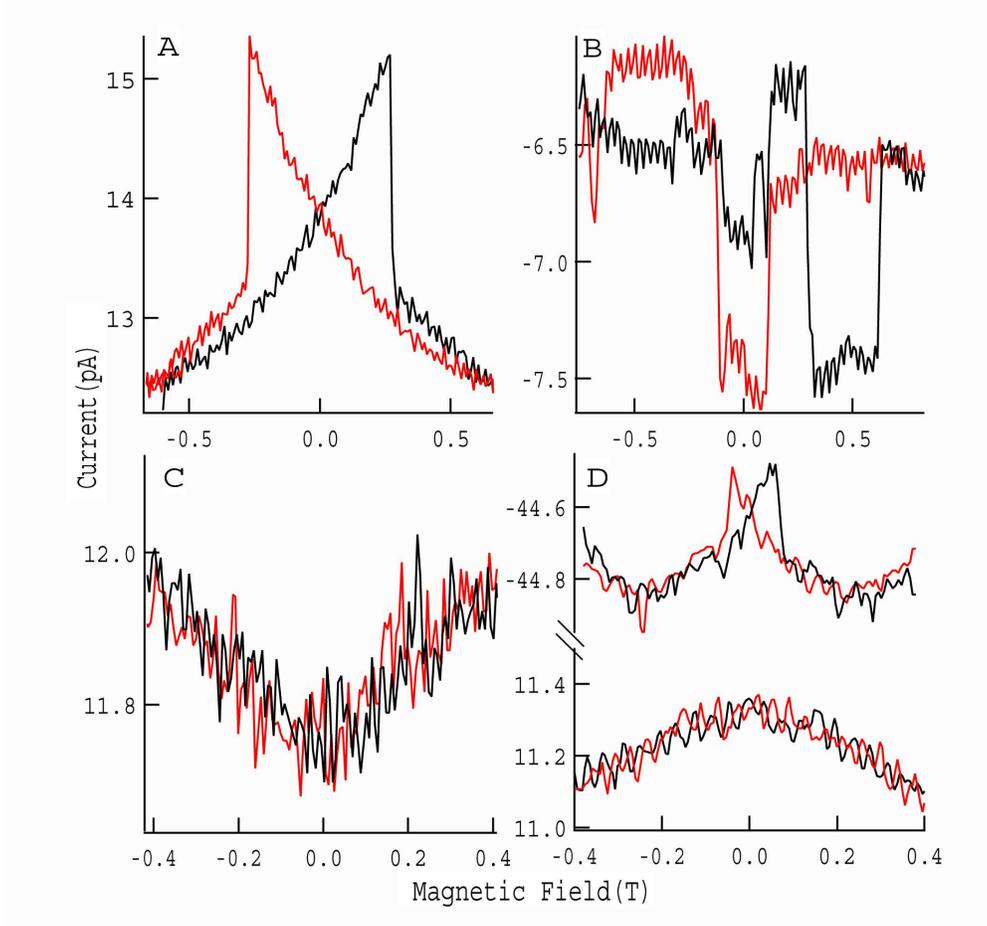}
    \caption{Examples of tunneling current versus magnetic field at
4.2K, for samples with A.Co, B.Fe, C.Ni, and D.Py nanoparticles.
In D, two pairs of curves correspond to two different samples.
Red(black) lines correspond to decreasing(increasing)
magnetic field. }
    \label{fig4}
\end{figure}
Current versus applied magnetic field (parallel to the film plane) is obtained by measuring the current at a fixed bias voltage while sweeping the magnetic field slowly.
Fig.~\ref{fig4}A-D show the representative magnetic hysteresis loops for Co, Fe, Ni, and Py samples at 4.2K, respectively.
As seen in Fig.~\ref{fig4}C and D, the Ni-nanoparticle and one Py nanoparticle have no magnetic hysteresis.
All the samples that lack hysteresis do so
at the lowest resolved tunneling current. The statistics of the presence of hysteresis for different materials are displayed in
Table. 1. All the Co (over 50) and Fe (6), one half of
the Py (out of 10), and only 2 of the 46 measured Ni
samples, with the switching field $0.05$T and $0.12T$, display magnetic hysteresis. The abrupt change in the hysteresis abundance between Co, Fe, Py, and Ni suggests a concept of minimum magnetic anisotropy necessary for magnetic hysteresis, akin to the concept of Mott's maximum metallic resistivity.
Note that the Ni
samples and the non-hysteretic Py samples still display
significant magneto-resistance. Among the hysteretic samples, the average magnetic switching field varies as Ni:Py:Co:Fe=$0.085$\,:$0.114$\,:$0.233$\,:$0.257$ (Tesla).

\begin{table*}
\caption{Hysteresis percentage versus magnetic anisotropy in different materials\cite{papaconstantopoulos, kittel}}%
\begin{ruledtabular}
\begin{tabular}{l |c |c |c |c |c | }
\hline
		Material & Hysteresis($\%$)   & $K_s$($\mu$eV/spin) & $K1$($\mu$eV/spin) &  $K2$($\mu$eV/spin) & $E_B$(meV) \\[0.5ex] \hline
	  Co\cite{paige}         & 100  & 105 & 64.1     & 8.3 &  $97\pm39$     \\ \hline
		Fe\cite{tung1982}      & 100  & 128  & 4.0    & 1.3 &   $51\pm20$   \\ \hline
		Py\cite{martinez2006}  & 50   &  67.1 & 0      & 0    &   $13\pm5$         \\ \hline
	  Ni\cite{gadsden}       & 4    & 37.9 &  -24.4  & 4.1  &    $7.7\pm3$ \\ \hline
\end{tabular}
\end{ruledtabular}
\end{table*}%

The magnetic energy $E_M(\vec{m})$ of the Fe/Ni/Py and Co nanoparticles can be written as $S_0[K_1(m_x^2m_y^2+m_y^2m_z^2+m_z^2m_x^2)+K_2m_x^2m_y^2m_z^2+K_s\vec{m}\hat{N}\vec{m}]$
and
$S_0[K_1m_z^2+K_2m_z^4+K_s\vec{m}\hat{N}\vec{m}]$, respectively, where
$S_0\hbar$ is the total spin in the nanoparticle, $\vec{m}$ is the magnetization unit vector, and $\hat{N}$ is the demagnetization tensor which is set to have 3 eigenvalues equal to 0.2, 0.3, and 0.5.
In the calculations, Euler angles between the principal axes of shape anisotropy and magnetocrystalline anisotropy axes are all equal to $\pi/5$. 
The results of the calculation of the energy barrier $E_B$ are shown in table 1, assuming the nanoparticle average volume obtained in Sec. 2. The error bar reflects the standard deviation in the nanoparticle volume. The abrupt change in magnetic hysteresis abundance is monotonic with the calculated $E_B$. Since 50\% of Py nanoparticles display hysteresis,
it follows that the minimum anisotropy energy barrier required for magnetic hysteresis in our samples at 4.2K is $\approx$13meV. 
This energy barrier is too large to explain our findings in terms of the reduction in the blocking temperature among these metals. 
Experimentally, the magnetometry on similarly-sized nanoparticles show that the blocking temperature varies between $13-30$K for 
Co and $6-20$K for Ni.~\cite{sarkar,luis,yoon,galvez,fonseca}
Though Co nanoparticles appear to have higher blocking temperature than Ni nanoparticles, the difference in blocking temperature is not sufficient to explain the vast contrast in the hysteresis abundance in our nanoparticles under electron transport. Theoretically, the Arrhenius flipping rate can be estimated as $\nu_{at}\exp(-E_B/k_BT)$. Assuming $\nu_{at}=10^{10}$Hz,
we obtain the magnetization flipping time of 100 hours, much longer than the time it takes to measure a magnetic hysteresis loop. We conclude that the breakdown in magnetic hysteresis we observe reflects the effect of sequential electron tunneling through the nanoparticle on magnetization. This effect will be discussed in Sec. 4.

\section{Numerical Simulations}

The magnetic Hamiltonian for Ni and Fe nanoparticles can be written as
\begin{eqnarray*}
\mathcal{H}(n,\vec{S_n})&=S_0[\frac{K_{1,n}}{2}(\alpha^2\beta^2+\beta^2\alpha^2+\alpha^2\gamma^2+\gamma^2\alpha^2+\beta^2\gamma^2+\gamma^2\beta^2)+\frac{K_{2,n}}{6}(\alpha^2\beta^2\gamma^2\\
&+\alpha^2\gamma^2\beta^2+\beta^2\alpha^2\gamma^2+\beta^2\gamma^2\alpha^2+\gamma^2\beta^2\alpha^2+\gamma^2\alpha^2\beta^2)]-K_s\vec{S_n}\hat{N}\vec{S_n}
\end{eqnarray*}

Co has uniaxial magnetocrystalline anisotropy, thus the magnetic Hamiltonian is
\begin{eqnarray*}
\mathcal{H}(n,\vec{S_n}) &=-S_0[K_{1,n}\gamma^2+K_{2,n}\gamma^4]-K_s\vec{S_n}\hat{N}\vec{S_n}
\end{eqnarray*}

Here, $n$ is the number of electrons in the nanoparticle while $S_0\hbar$ is the total spin.  $K_{1,n}$, $K_{2,n}$, and $K_s$ represent magnetocrystalline anisotropy constants and the shape anisotropy constant per spin, respectively. $\alpha$,$\beta$,$\gamma$=$S_x$,$S_y$,$S_z/S_0$. $\hat{N}$ is the demagnetization tensor.
The magnetocrystalline anisotropy constants per unit volume  $K_{1,V}$ and $K_{2,V}$ are obtained from Refs. \onlinecite{tung1982,paige,martinez2006,gadsden} . We obtain $2S_0/N_a$ from Ref.~\cite{papaconstantopoulos}, where   $N_a$ is the total number of atoms in the nanoparticle. Then,  $S_0/V=\rho S_0/N_aM_A$, where $\rho$ is the mass density  and $M_A$ is the atomic mass. The average value of the magnetocrystalline anisotropy constants per spin $(K_{1},K_{2})$ are $(K_{1,V},K_{2,V})V/S_0$, respectively, while, $K_s=\mu_0M_s^2V/2S_0$, where $M_s$ is the saturation magnetization obtained in Ref. \onlinecite{kittel,martinez2006}.
Because of spin-orbit anisotropy fluctuations, ($K_{1,n},K_{2,n}$) fluctuate around $(K_{1},K_{2})$, respectively, according to the number of electrons $n$. We expect that the average magnetic anisotropy has strong material dependence, while the mesoscopic fluctuations in the total magnetic anisotropy energy due to single electron tunneling on/off, are independent of the material.~\cite{cehovin,usaj,brouwer3} In Co nanoparticles, the change in total magnetic energy after electron tunneling-on is $S_0\Delta K\sim0.4$meV.\cite{usaj} In Ni nanoparticles of the same size at 4.2K or below, $S_0$ and $K$ are both  $\approx 1/3$ of the values in Co nanoparticles. So, the relative fluctuations in magnetic anisotropy $\Delta K/K$ in Ni nanoparticles are enhanced $9-10$ times compared to Co nanoparticles with the same size. At the same time, the magnetic energy barrier in Ni nanoparticles is suppressed because of the cubic symmetry and weak shape anisotropy. As a result, the magnetization in Ni nanoparticles is significantly more susceptible to perturbation by electron transport compared to Co or Fe nanoparticles.

The numerical simulation of the magnetization dynamics is based on the master equation modified from that in Ref.~\cite{parcollet}. In the sequential tunneling regime, the number of electrons in the nanoparticle  hops between $n$ and $n+1$. Assuming the electron tunnels through only one  minority single electron state $j$, which reduces the spin of the nanoparticle by $1/2$ after the electron tunneling-on event  (including more electron states does not affect the result in a major way), the master equation can be written as


\begin{equation} \label{master}
\begin{split}
\frac{\partial P_{\alpha}}{\partial t}=&\sum_{\alpha'}\sum_{l=L,R}\sum_{\sigma=\uparrow,\downarrow}\Gamma_{l\sigma}\left\{\left|<\alpha'|c_{j\sigma}|\alpha> \right|^{2}\left[-(1-f_l(\Delta E_M))P_{\alpha}+f_l(\Delta E_M)P_{\alpha'}\right]+
\right. \\ & \left.
\lvert<\alpha'|c^{\dagger}_{j\sigma}|\alpha>\rvert^{2}\left[-f_l(-\Delta E_M)P_{\alpha}+(1-f_l(-\Delta E_M))P_{\alpha'}\right] \right\}+\\
& \Gamma_B\sum_{\alpha',\epsilon=\pm}\epsilon\left|<\alpha'|S_x-i\epsilon S_y|\alpha> \right|^{2}\left[\rho_B(\epsilon\Delta E_M)n_B(\Delta E_M)P_{\alpha'}+\rho_B(\epsilon\Delta E_M)n_B(-\Delta E_M)P_{\alpha}\right]
\end{split}
\end{equation}

The first part of the master equation describes the magnetic tunneling transitions. Here, $|\alpha>$ and $|\alpha'>$ represent magnetic eigenstates of the nanoparticle , i.e., the eigenstate of $\mathcal{H}(n,\vec{S_n})$ with $n$ or $n+1$ electrons.  $|\alpha>$ and $|\alpha'>$  can be obtained as superpositions of the eigenstates of $\hat{S}^2$ and $\hat{S_z}$, which are the pure spin-states $|S_0,m>$.  $c_{j\sigma}$ ($c^{\dagger}_{j\sigma}$) is the annihilation (creation) operator for an electron with spin $\sigma$ on level $j$. $\Gamma_{l\sigma}$ denotes the tunneling rate to level $j$ through the leads $l=L,R$ for electron with spin $\sigma$ and $f_l$ is the Fermi distribution in the leads. $\Delta E_M=E_{\alpha}-E_{\alpha'}$.   $<\alpha'|c^{\dagger}_{j\sigma}|\alpha>$ is the tunneling transition matrix element for transition between the initial state $|\alpha>$ and the final state $|\alpha'>$.  The second part of the master equation describes the magnetic damping due the coupling to the bosonic bath. $\Gamma_B$ is the rate related to the damping rate $1/T_1$.  $\rho_B(\Delta E_M)$ is the spectral density of the boson  which is set to be constant because it varies very slowly\cite{parcollet}. $n_B$ is the Bose-distribution function.

\begin{figure}
\centering
\renewcommand{\thefigure}{5}
\includegraphics[width=0.95\textwidth]{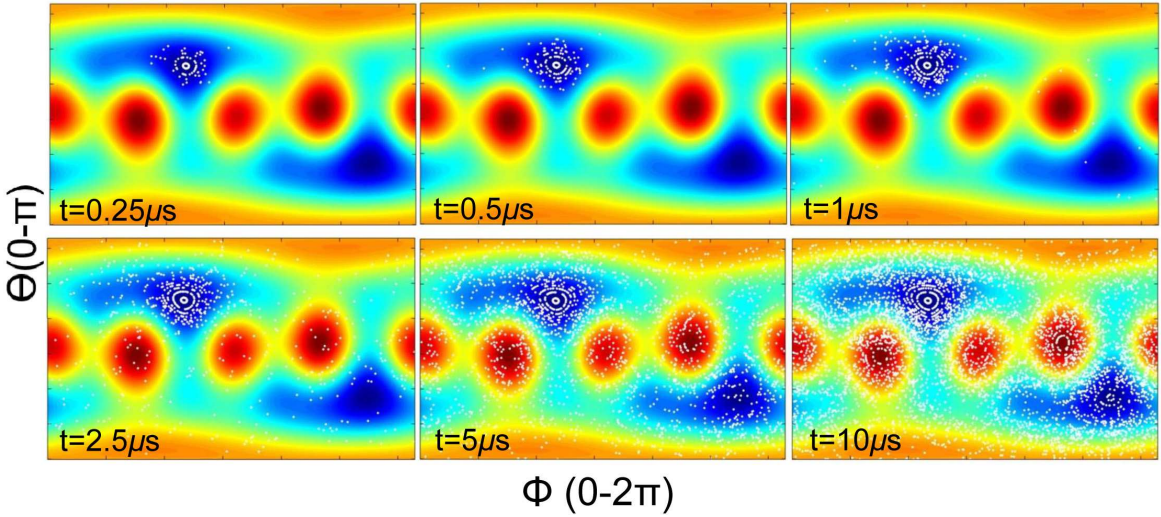}
\caption{Simulations of the time evolution of the magnetization vector in a Ni nanoparticle. The blue-yellow-red scale indicates the magnetic anisotropy energy landscape from low to high versus the polar angle  $\Theta$ and the azimuthal angle $\Phi$ in a Ni nanoparticle. The value of t is the past time counted from the beginning of electron tunneling.  The white dots signify possible angle positions at the specified time for each simulation. As time progresses, the likely magnetization position is spread out over phase space and approaches a random isotropic distribution. }
\end{figure}

Fig. 5 shows the motion of the magnetization statistical distribution versus time. The magnetic anisotropy energy vs ($\Theta,\Phi$) represents the relation between magnetic energy and magnetization direction, where $\Theta, \Phi$ are the two polar angles. Each magnetic state corresponds to a contour on the magnetic anisotropy energy landscape. Classically, one would consider the magnetic state as the magnetization precessing on the contour corresponding to that state. A random dot is selected on a contour to indicate the presence of the magnetic state related to that contour. The number of dots on each contour is decided by the probability of the corresponding magnetic state. The plotted dots together represent the distribution of the magnetization over $\Theta,\Phi$-space at that time.

In the simulation for Fig.~5, $S_0=100$. $\Gamma_{L,R\sigma}=6\times10^7$ which corresponds to a current about $5$pA. $\Gamma_B=2\times10^3$ leads to a relaxation time of $\sim2.5\mu$s. $(K_{1,n},K_{2,n})=1.25(K_{1},K_{2})$ and $(K_{1,n+1},K_{2,n+1})=0.75(K_{1},K_{2})$.  $K_s$ is
taken from Table.~1. $\hat{N}$ is set to have 3 eigenvalues equal to 0.2, 0.3,
and 0.5. 
Varying $\hat{N}$ and the Euler angles, which were earlier defined, does not affect the qualitative result. The magnetic field is set to 0.001T to eliminate Kramers degeneracy.  We set the nanoparticle to be initially at the ground state with $n$ electrons. Then we iterate the master equation for 40$\mu$s. The magnetic state distribution of the nanoparticle gradually spreads from the ground state to other states and eventually becomes isotropic as shown in Fig.~5.

\section{Discussion and Conclusions}

The transfer of a single electron into the magnetic nanoparticle creates a fluctuation in the spin-orbit energy of the nanoparticle.~\cite{usaj,kleff,cehovin,brouwer3} Such a fluctuation in turn creates a spin-orbit torque that is exerted on the magnetization. In the previous section, we show how  a fluctuating spin-orbit torque can lead to isotropic distribution of the magnetization. The fluctuating spin-orbit torques are mesoscopic effects and do not depend significantly on the material of the nanoparticle.~\cite{cehovin,brouwer3} But, the nonfluctuating magnetic anisotropy, such as magnetocrystalline and magnetostatic shape anisotropy, depends strongly on the material. As the magnetic anisotropy of the nanoparticle is reduced, the  strength of the fluctuating spin-orbit torques relative to the deterministic torques will increase, creating a noise floor which sets the limit on magnetic anisotropy below which magnetic hysteresis cannot be observed in sequential electron tunneling.
Such a noise floor is reflected by the abrupt change in magnetic hysteresis abundance in similarly sized Ni, Py, Fe, and Co nanoparticles. By contrast, in thermal equilibrium, magnetometery of the ensembles of similarly sized Co and Ni nanoparticles show much less dramatic change in the blocking temperature.~\cite{sarkar,luis,yoon,galvez,fonseca} The minimum magnetic anisotropy also explains prior measurements of voltage biased single magnetic molecules, in a double tunneling barrier, which showed no signs of magnetic hysteresis, even at temperatures much lower than the blocking temperature,~\cite{burzuri,moon} notwithstanding that the magnetometry of ensembles of such molecules showed magnetic hysteresis below the blocking temperature.~\cite{thomas,friedman}

It may be surprising to the reader that the minimum magnetic anisotropy is found to be independent of the size of the tunneling current through the nanoparticle. The tunneling current we use in the measurements of current versus magnetic field, varies between 1pA and 100pA. A Co nanoparticle at 100pA is likely to exhibit magnetic hysteresis, while a Ni nanoparticle at 1pA is highly unlikely to do so. The ratio of the applied tunneling currents is one order of magnitude larger than the ratio of the energy barriers between the average Co and Ni nanoparticle. Since the data presented in this paper were gathered, we have studded single Ni nanoparticles at mK-temperature, and discovered that $2$ out of the  $5$ measured Ni nanoparticles display magnetic hysteresis at the onset voltage for sequential electron tunneling.~\cite{gartland2} The hysteresis in current versus magnetic field was abruptly suppressed in the bias voltage range starting just above the lowest discrete energy level for single-electron tunneling. The abrupt suppression of magnetic hysteresis versus bias voltage was explained in terms of the magnetization blockade, which was caused by the bias voltage dependent damping rate.~\cite{waintal,gartland2} In the magnetization blockade regime, the ordinary spin-transfer~\cite{waintal} or the spin-orbit torques~\cite{gartland2} are damped because of the small bias energy available for single-electron tunneling. In the voltage region where the magnetization is blocked, the spin-transfer and damping rates are both proportional to the electron tunneling rate, regardless of which spin-transfer mechanism is at play (e.g. the ordinary  spin-transfer or the spin-orbit torques). A change in the bias-voltage will change the damping rate,~\cite{gartland2} but the effective magnetic temperature, controlled by the ratio of the damping rate and the spin-transfer rate,~\cite{waintal} will be independent of the electron tunneling rate. It is reasonable to assume that the magnetic nanoparticle will exhibit magnetic hysteresis in our experimental times scales, if the flipping time given by the Arrhenius law, based on the attempt frequency and the ratio of the energy barrier and the effective magnetic temperature, is longer than the hysteresis measurement time.
Since neither the energy barrier nor the effective magnetic temperature depend on the tunneling current, this explains, at least in principle, why magnetic hysteresis abundances in our samples are so weakly dependent on the tunneling current. The characteristic temperature above which the two hysteretic Ni nanoparticles stop displaying magnetic hysteresis is $2-3$K.~\cite{gartland2} That characteristic temperature corresponds to the magnetization blockade energy, which is comparable to the single-electron anisotropy. We can conclude that the minimum magnetic anisotropy is the limiting anisotropy of the nanoparticle below which magnetic hysteresis cannot be guaranteed. If the magnetic hysteresis does occur in the nanoparticle with magnetic anisotropy smaller than the minimum magnetic anisotropy, it will do so below 2-3K temperature, the characteristic temperature of the magnetization blockade.


In summary, we have performed magnetoresistance measurements on a variety of ferromagnetic materials 1-5 nm in diameter at 4.2K, and found an abrupt change in magnetic hysteresis abundance between Ni, Py, Fe, and Co nanoparticles. This abruptness leads to the conclusion that there is a minimum magnetic anisotropy energy in a metallic ferromagnetic nanoparticle or a magnetic molecule out of equilibrium, required to guarantee magnetic hysteresis at low temperatures. The size of the tunneling current does not affect the minimum magnetic anisotropy. Our finding has an implication for the miniaturization of magnetic random access memory. It demonstrates a limit below which reliable reading of soft-layer magnetization cannot be predictable below the blocking temperature. Research supported by the U.S. Department of Energy, Office of Basic Energy Sciences, Division of Materials Sciences and Engineering under Award DE-FG02-06ER46281. We thank Dr. Ding from the Microscopy Center, School of Materials Science and Engineering at Georgia Institute of Technology for his help in taking the TEM image of the Ni nanoparticles.

\bibliography{career1}

\end{document}